\documentclass[conference, ]{IEEEtran}
\IEEEoverridecommandlockouts
\usepackage{cite}
\usepackage{amsmath,amssymb,amsfonts}
\usepackage{algorithmic}
\usepackage{graphicx}
\usepackage{textcomp}
\usepackage{xcolor}
\usepackage{url}
\usepackage{soul}
\usepackage{float}
\usepackage{url}
\usepackage{booktabs}
\usepackage{tabularx}
\usepackage{makecell}

\usepackage{subfig}
\usepackage{graphicx}
\usepackage{etoolbox}
\usepackage{enumitem}
\usepackage{booktabs}

\usepackage[font=small, labelsep=period]{caption}
\usepackage[hidelinks]{hyperref}
\usepackage[table]{xcolor}
\hypersetup{
    colorlinks=true,
    urlcolor=blue,
    citecolor=black,
    linkcolor=black
}

\newcommand{\osf}{{\url{https://osf.io/82wxe}}}

\newcommand{\etal}{et al.}
\newcommand{\etals}{et al.'s}

\newcommand{\eg}{{e.g.,}}

\newcommand{\secref}[1]{\hyperref[#1]{Sec.~\ref*{#1}}}
\newcommand{\appendixref}[1]{\hyperref[#1]{Appendix~\ref*{#1}}}
\newcommand{\figref}[1]{\hyperref[#1]{Fig.~\ref*{#1}}}
\newcommand{\eqnref}[1]{\hyperref[#1]{Eqn.~\ref*{#1}}}
\newcommand{\tabref}[1]{\hyperref[#1]{Table ~\ref*{#1}}}

\newcommand{\hlc}[2][yellow]{{%
      \colorlet{foo}{#1}%
      \sethlcolor{foo}\hl{#2}}%
}
\definecolor{quoteColor}{HTML}{ff5733}
\newcommand\qt[1]{\hlc[quoteColor!10]{``#1''}}
\definecolor{exampleColor}{HTML}{9370DB}

\newcommand{\pxx}[1]
{\textbf{P}$_{\textrm{#1}}$}

\newcommand{\pai}[1]{P$_{\textbf{AI#1}}$}
\newcommand{\plb}[1]{P$_{\textbf{NB#1}}$}
\newcommand{\por}[1]{P$_{\textbf{OR#1}}$}
\newcommand{\pxl}[1]{P$_{\textbf{XL#1}}$}

\newcommand{\parahead}[1]
{%
  \paraheadd{#1}.
}

\newcommand{\paraheadd}[1]
{%
  \vspace{0.07in}%
  \noindent%
  \textbf{\textit{#1}}%
}

\def\subsubsec#1
{\subsubsection{#1}}

\def\BibTeX{{\rm B\kern-.05em{\sc i\kern-.025em b}\kern-.08em
    T\kern-.1667em\lower.7ex\hbox{E}\kern-.125emX}}
\begin{document}


\newcommand{\insertfig}{
\vspace{1.5em}\includegraphics[width=\linewidth]{figures/dimensions-cut.pdf}\captionof{figure}{
    For each TDoPs~\cite{jakubovicEP23} dimensional cluster, we observe trade-offs that data wrangling tools approach in different ways. 
        While we place the tools in these spaces, their placement is less valuable than the spaces that they occupy, as special cases may change that positioning.
        We forgo Adoptability because our study focuses on short-term usage by practitioners, which limits commentary on factors like learnability.  }\addtocounter{figure}{-1}\label{fig:teaser}
\vspace{-1em}}
\makeatletter
\apptocmd{\@maketitle}{\vspace{-3em}}{}{}
\apptocmd{\@maketitle}{\centering\insertfig}{}{}
\makeatother


\title{How Wrangling Tools Shape Wrangling: \\A Technical Dimensions Analysis}


\author{\IEEEauthorblockN{1\textsuperscript{st} Shiyi He}
\IEEEauthorblockA{\textit{University of Utah} \\
Salt Lake City, Utah 
}
\and
\IEEEauthorblockN{2\textsuperscript{nd} El Kindi Rezig}
\IEEEauthorblockA{\textit{University of Utah} \\
Salt Lake City, Utah 
}
\and
\IEEEauthorblockN{3\textsuperscript{rd} Paul Rosen}
\IEEEauthorblockA{\textit{University of Utah} \\
Salt Lake City, Utah 
}
\and
\IEEEauthorblockN{4\textsuperscript{th} Andrew M. McNutt}
\IEEEauthorblockA{\textit{University of Utah} \\
Salt Lake City, Utah 
}
}

\maketitle

\begin{abstract}
Wrangling consumes a disproportionate share of the effort associated with any data project. 
While a variety of tools support it, relatively little is known about how their differing interface forms shape the way people actually wrangle.
We conduct a between-subjects (N=40) observational study of data cleaning tasks performed in tools spanning distinct interface paradigms: Jupyter (notebook), Excel (spreadsheet), ChatGPT (conversational AI), and OpenRefine (visual wranglers).
We situate our observations within the Technical Dimensions of Programming Systems framework, which we use as a conceptual scaffold for comparing across interface paradigms.
Within the context of our study, the results suggest that tool affordances steer user strategies but do not determine outcomes. There is no consistent advantage of any single tool, nor convergence of results within tools observed across our outcome measures.
Instead, we identify trade-offs and connect them with observed practice.
For example, a key tension is between data- and abstraction-centered interfaces, where data-centered interfaces encourage opportunistic cleaning rather than systematic, planned transformations found in abstraction-focused tools (but come with a cognitive burden).
Tool design, beyond mere functionality, plays a structuring role in how data work unfolds.

\end{abstract}

\begin{IEEEkeywords}
Data wrangling,
interface paradigms,
user study
\end{IEEEkeywords}

\addtocounter{figure}{1}

\section{Introduction}

Data wrangling is ubiquitous: wherever data is used, it must first be brought into a form that facilitates its use~\cite{Hellerstein2018SelfServiceDP}.
This process involves preparing data for subsequent usage, which can involve numerous manipulation operations such as cleaning,
transformation,
and integration~\cite{kandel2011wrangler}.
It is carried out by many types of users, including analysts, data journalists, and engineers.
It can be challenging: demanding iterative refinement~\cite{krishnan16cleaning} and substantial domain expertise, sometimes necessitating as much as 80\% of the effort in data projects~\cite{kandel2011research}.
Wrangling occurs in a range of contexts including everyday substrates (such as notebooks or spreadsheets), emerging environments (like natural language UIs), and dedicated wrangling tools (as exemplified by visual wranglers like Trifacta).

Despite its prominence, it is not clear how differences in interaction paradigms affect how people approach wrangling and the quality of their results.
To bridge this gap,
we conducted a (N=40) crowd-work observational study in which participants conducted a pair of data wrangling tasks using one of several representative tools, chosen to span a range of interaction modalities---including notebooks  (Jupyter~\cite{jupyterLite}), spreadsheets (Excel), natural-language UIs (ChatGPT), and dedicated wrangling tools (OpenRefine~\cite{verborgh2013using}).
We quantitatively investigate \emph{how well} (\secref{sec:quant})  these tools support wrangling as well as \emph{how} (\secref{sec:tdops}) they shape it, using Jakubovic \etals{}~\cite{jakubovicEP23} Technical Dimensions of Programming Systems (TDoPS).

We find that, instead of a single best tool, each design embodies a distinct set of trade-offs (see \figref{fig:teaser}) that shape problem formulation, exploration, and solution development.
Tools like Excel center data in their interface (inviting interaction with visible errors), whereas others, such as Jupyter, abstract it (pushing towards more systemic but higher complexity cleaning).
Errors in these systems arise from both usage (often mediated by operational complexity) and the data itself, each requiring different design responses.
A persistent tension runs between general-purpose tools used for wrangling (which obscure or abstract domain-specific affordances) and dedicated wranglers (which gain task fit at the cost of generality, often forcing users to coordinate multiple tools).
These choices shape how wrangling is performed and conceptualized, with characteristic failures emerging when paradigms are poorly matched to task.
Wrangler design is not solely technical, but also relates to how users think and reason with these inherently human-centric~\cite{s2024cognitive} systems.

By surfacing the trade-offs that underlie interface paradigms, this work provides a new perspective on how design choices shape data wrangling in practice, beyond mere functionality.
Wrangling is essential to data work, and the design choices that shape it demand careful consideration.

\section{Related Work}

Many tools have been developed for data wrangling.
Wrangler~\cite{kandel2011wrangler, wrangler_cidr} was an early standalone tool to this effect (preceded by Potter's wheel~\cite{potterwheel}) that centered on graphical explanation of transformations, mediated through a spreadsheet-like interface. This was subsequently commercialized as Trifacta.
A variety of commercial tools have since emerged, including Tableau Prep~\cite{tableauprep}, Alteryx~\cite{alteryx}, KNIME~\cite{knime}, and Talend~\cite{talend}, among a range of others.
Profiler~\cite{kandel2012profiler} expanded on Wrangler with means for visually inspecting data quality issues.
Shrestha \etal{}~\cite{shrestha2021unravel, shrestha2023detangler} explored augmenting wrangling code with visualizations to support understanding and debugging.
Data noodles~\cite{gorinova2016transforming} explored spreadsheet specific wrangling affordances via demonstration, which Dango~\cite{chen2025dango} subsequently integrated with natural language.
Buckaroo~\cite{rezig26buckaroo} explored using direct manipulation of errors in a visual wrangler as a means to select parts of the data to be cleaned---a strategy initially explored in Arachnid~\cite{shou2019arachnid}.
This body of work illustrates the breadth of approaches in the area.
While we do not study these systems, our findings characterize patterns that likely generalize to these classes of wrangling tools.

More closely related to this paper are works exploring the human factors of data wrangling.
Bhowmick \etal{}~\cite{s2024cognitive} highlighted that data systems should be informed by cognitive psychology to better align with users' thinking processes.
Kasica \etal{} characterized~\cite{kasica2023dirty} domain-specific challenges journalists face in data preparation and proposed~\cite{kasica2020table} a framework for multi-table wrangling.
Kandel \etal{}~\cite{kandel2012enterprise} interviewed industrial data analysts about different roles through which users approach data analysis, finding key challenges in wrangling include semi-structured data, integration, and advanced analytics.
Our findings complement these in that we observe that these challenges are both facilitated and complicated by interface design choices.
Coscia \etal{}~\cite{coscia2023preliminary} explore combining data integration processes in visual analytics systems, considering integrated versus combined interfaces---an issue we examine as well.
Gorinova \etal{}~\cite{gorinova2016end} framed data wrangling as an end-user programming problem, which informs our use of TDoPs. Whereas they recommend table-based demonstration-centered interactions, we highlight that this focus can come at the cost of more procedural or abstracted thinking.
Krishnan \etal{}~\cite{krishnan16cleaning} surveyed data analysts and infrastructure engineers about data cleaning practices, finding that cleaning is an iterative and ad hoc process---which we also observe.
Similarly, Sultanum \etal{}~\cite{sultanum2024data} conducted an interview study to understand how trust is mediated within data products, highlighting how validation and verification interact.
Petricek \etal{}~\cite{petricek2022ai} explored AI agents as data wrangling tools, a consideration we echo by studying ChatGPT.
In analyzing their tool Dango, Chen \etal{}~\cite{chen2025dango} found that mixed-initiative interactions can improve confidence in LLM-based wrangling.
While they study feature variations in their tool,
we consider several in-the-wild tools.

\begin{figure}[t]
    \centering
    \includegraphics[width=\linewidth]{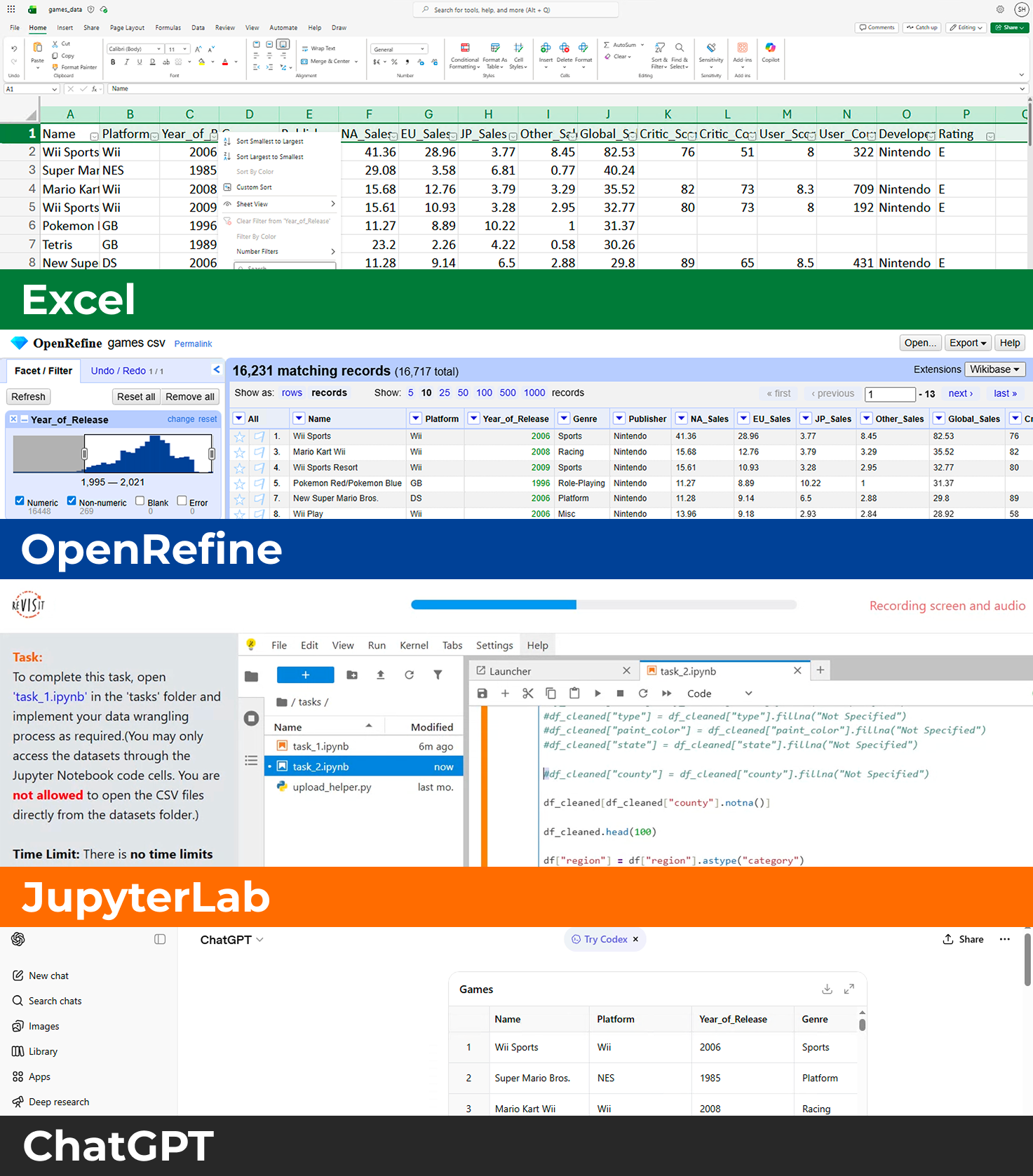}
    \caption{
Our study was deployed using ReVISit~\cite{cutler20252025ReVISit}, with a tool-specific environment embedded for each condition.}
    \label{fig:all-tool}
    \vspace{-2em}
\end{figure}

\section{Methods}

We conducted a (N=40) between-subjects observational study examining how current tools support data wrangling. We consider \emph{how well} tools support wrangling, which we consider via a quantitative analysis (\secref{sec:quant}), and \emph{how} they shape user strategies, which we address qualitatively via TDoPS (\secref{sec:tdops}).
This study was marked exempt by our institution's IRB.

Participants were assigned a single tool to use for the duration of the study.
We selected a set of tools that varied by paradigm:
spreadsheet (Excel), visual wrangler (OpenRefine), computational notebook (Jupyter), and natural-language chat (ChatGPT).
These tools differ in the interaction style and capabilities, ranging from domain-specific to more compositional and open-ended workflows.
Despite these differences, all tools support the capabilities required by the study.
There are others that would be useful to look at, but are beyond the scope of this study.
Most notably, we considered Trifacta (which has since been absorbed into Alteryx), however it was prohibitively expensive to attain a version of Alteryx that involved the core parts of Trifacta.
Research tools~\cite{chen2025dango, xiong2022revealing}
might be useful to study, but  we focus on mature, paradigmatically representative tools (\eg{} OpenRefine has $\sim$12k GitHub stars).

\begin{figure}[t]
    \centering
    \includegraphics[width=\linewidth]{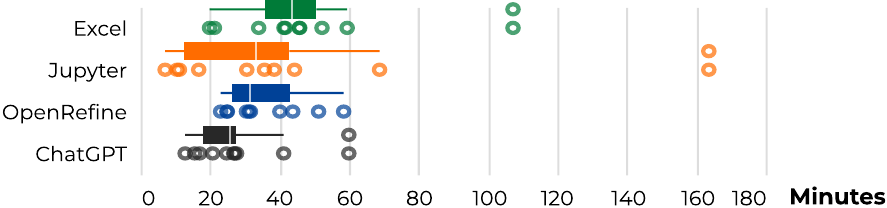}
    \caption{Completion time by condition.
    }
    \label{fig:timing}
    \vspace{-2em}
\end{figure}

Participants completed a pair of representative data wrangling tasks.
In the first, they wrangled a Game Rating Dataset~\cite{game}, requiring them to filter, standardize, and remove irregularities (\eg{} type errors, outliers).
In the second, more open-ended task, they wrangled a Used Cars Dataset~\cite{car}, seeking to
standardize data types, clean anomalies, and remove irrelevant columns. The first task had a 10-minute soft time limit, while the second was untimed.
See appendix for task specifics.
We used these tasks because they are representative, while being understandable without domain knowledge, and featured both constrained and open wrangling processes respectively.
As real-world wrangling is complex and varied, these tasks necessarily cover only a subset of lived practices.
Accordingly, we stress the exploratory nature of our work.
Following consent, participants were asked to think aloud as they worked through the tasks.
After the primary study tasks were completed, participants were prompted to answer a series of questions aloud about their experiences using the tools, and then completed experience surveys, including the NASA-TLX and UMUX.
Participants took an average of $38\pm27$ minutes overall, although this varied by tool, as in \figref{fig:timing}.

We recruited 40 participants from the freelance platforms Upwork (39) and Fiverr (1).
We targeted participants who were familiar with data wrangling and had prior experience using each tool through tool-specific advertisements.
We initially used Prolific for recruitment. However, no pilot participants had sufficient wrangling or coding experience to complete the study.
We shifted to the freelancing platforms Fiverr and Upwork.
We piloted 5 freelancers, 4 of whom were used in our data analysis as their experiment was identical to the final study.
Due to data issues or not following instructions, 5 participants were dropped (but paid).
Participants and pilots were paid \$40.
They had $5.5 \pm 4.5$ years of experience with data wrangling and rated themselves as very familiar with it ($5.7\pm1.2$ out of 7).
Participants are \pai{x}, \plb{x}, \por{x}, \pxl{x} for ChatGPT, Jupyter, OpenRefine, and Excel.

Our study was implemented in ReVISit~\cite{cutler20252025ReVISit}, which afforded capture of think-aloud audio and screen-capture (\figref{fig:all-tool}).
To support remote deployment, consistent instrumentation, and constraints around licensing and installation, participants accessed the study tools through browser-based configurations.
For Excel (web version), ChatGPT (5.2), and OpenRefine (3.10.0), participants connected to a Windows VM via noVNC~\cite{noVNC} in an iframe. For the Jupyter condition, participants used JupyterLite~\cite{jupyterLite}, a JupyterLab distribution that runs entirely in the browser and supports the relevant packages for our tasks, including pandas.
We used JupyterLite over a hosted JupyterLab instance because it runs without a backend, eliminating the need to provision and secure a Python environment per participant.
For consistency, we use Jupyter to refer to this condition throughout the paper.
In piloting, we found the VNC connection added no appreciable lag.
See the Jupyter condition with participant replays at \href{https://wrangler-tdops-study.netlify.app/a-notebook}{https://wrangler-tdops-study.netlify.app/a-notebook}.

\section{Quantitative Analysis}
\label{sec:quant}

First, we consider \emph{how well} participants did per tool and task.
Unlike other areas, where results might be analyzed through a comparison with ground truth, there are many ways to complete these tasks.
To account for this variance, we generated a set of plausible, task-aligned cleaning approaches to compare participant outputs against these pre-defined baselines (see appendix for the generation details). We refer to these as silver tables to distinguish them from a single idealized gold standard. These silver tables do not represent absolute quality, instead, they provide a reference space of reasonable solutions that satisfy the task requirements in various ways. While this approach has limitations, it offers a practical baseline for assessing the relative quality of participant outputs.
As our work is exploratory and our sample size is limited, we do not make statistical inferences or claims of performance differences.

We computed pairwise differences among all participant results, the silver tables, and the original dataset using Jaccard similarity, yielding  \figref{fig:quant}.
We use Jaccard similarity because it is interpretable, symmetric, and robust to the differences in row ordering. To ensure that the resulting clusters reflected meaningful wrangling patterns, we further examined how outputs within each cluster were similar, identifying their shared cleaning strategies through participants' video and think-aloud data.
For Task~2, we canonicalized the columns via a manual alignment, including inserting columns of nulls for those that had been deleted and enforcing a consistent ordering.
We observe that differences in paradigms do not
influence wrangling success consistently and that task context constrains the space of viable solutions.

\parahead{Tools alone do not drive successful outcomes}
Only a few participants from each tool generated outputs close to the silver tables in task 1 (\eg{} cluster 2), indicating that the tools alone do not reliably lead to successful outcomes even for the relatively simple Task~1.
Moreover, some clusters (\eg{} clusters 1 and 6) include results from multiple tools, suggesting that similar transformations can be achieved across different interfaces.
For example, participants across all tool conditions handled type-mismatched values in numeric columns in similar ways, such as by dropping affected rows or replacing them with nulls.
While each tool provides its own mechanisms for performing these transformations, they ultimately support a comparable set of common data-cleaning operations (echoing the collection of operations developed in Wrangler~\cite{kandel2011wrangler}).
This suggests effective data wrangling depends not only on tool capability, but may also on other factors such as user skill, domain knowledge, and external resources.

\parahead{Task constrains the space of viable strategies}
The clustering patterns suggest that the strategies used are more constrained in Task~1 than in Task~2, as reflected by the larger, cohesive clusters on the left.
This is likely due to the more explicit and restricted requirements of Task~1, which limit the space of viable solutions and lead to more similar approaches.
In contrast, Task~2 is more open-ended, allowing greater flexibility and resulting in substantial divergence both across tools and within individuals.
For example, in Task~1, cluster 3 includes a higher proportion of ChatGPT participants, whereas this pattern does not appear in Task~2.
This suggests that AI tools may be more consistent in supporting well-specified, constrained tasks, echoing prior findings that AI-generated code tends to perform better with explicitly instructed wrangling tasks such as transformation\cite{li2024towards}. However, it struggles in open-ended data wrangling scenarios that require iterative exploration and contextual judgment. In more complex (and realistic) data wrangling contexts, these affordances would not seem to be sufficiently supportive alone, leaving much of the burden of adapting to task complexity to users.

\section{TDoPS Analysis}
\label{sec:tdops}

Next, we consider \emph{how} participants did with each of the tools.
To do so, we analyze behavior exhibited during the study (\eg{} screen-captures, think-aloud data, and post-task questionnaires) through the lens of Technical Dimensions of Programming Systems (TDoPs)~\cite{jakubovicEP23} to examine how each tool environment shaped participants' wrangling practices (see \figref{fig:teaser}). This framework conceptually scaffolds analysis of the ways users interface with programming systems through 7 clusters of dimensions, which we step through each of in this section---with the exception of adoptability, consideration of which requires longer-term evidence (although we consider it in the appendix).
We conducted a deductive thematic analysis~\cite{clarke2017thematic}, wherein codes were drawn from TDoPs~\cite{jakubovicEP23}.
The primary coding was conducted by the first author (see appendix for coding details), with the results being iteratively refined with the rest of the team.
Different clusters have differing levels of salience, and corresponding levels of detail.
We follow earlier works that use TDoPs~\cite{horowitz2025sculpin, edwards2025baseline, cutler20252025ReVISit, mcnutt20232023Projectional}, or the closely related Cognitive Dimensions of Notation (CDN)~\cite{holwerda2018usability, jansen2019xlblocks, gridlets} for heuristic analysis of programming systems.

\subsection{Interaction: How do users manifest their ideas, evaluate the result, and generate new ideas in response?}
\label{sec:interaction}

Data wrangling involves iterative cycles~\cite{krishnan16cleaning} of transformation, evaluation, and approach refinement.
Effectiveness depends not only on available functionality, but also on how well tools support users in interpreting system feedback (evaluation) and translating intentions into correct actions (execution), per Norman~\cite{normanDesign}.

\begin{figure}[t!]
    \centering
    \includegraphics[width=\linewidth]{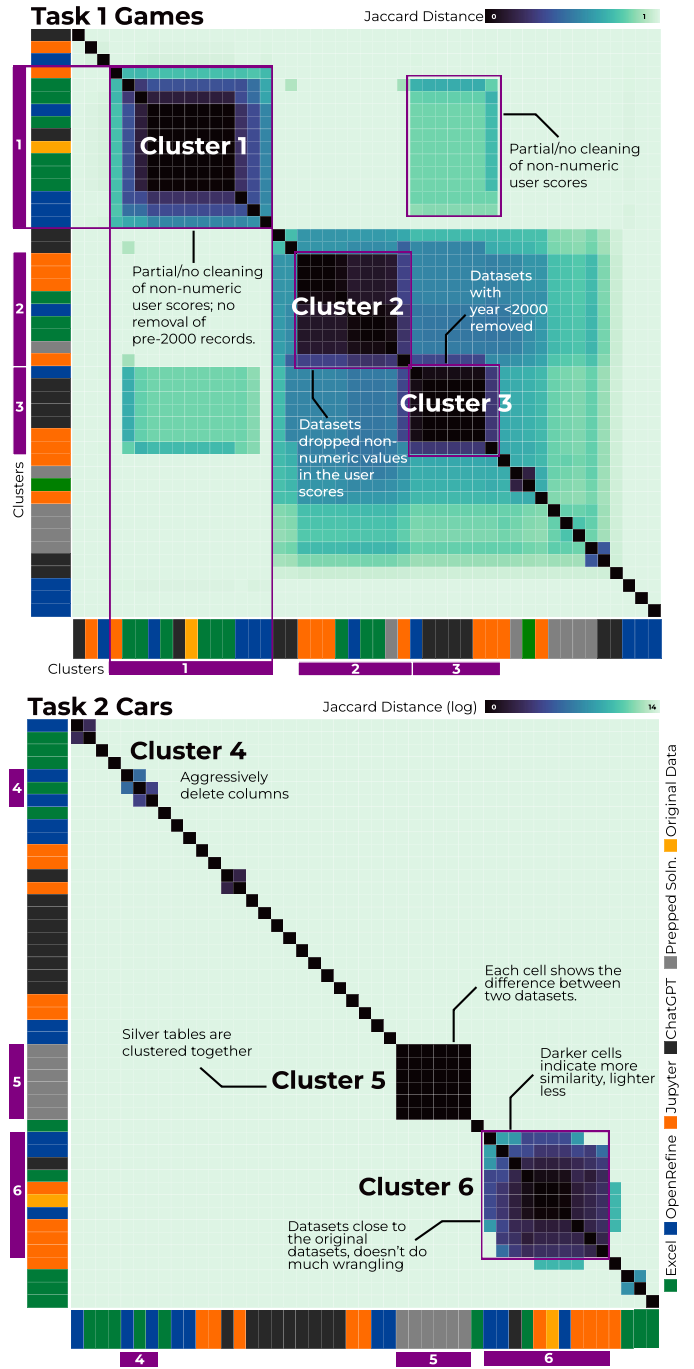}
    \caption{
        Similarity between participant outputs, the original dataset, and our silver tables, ordered by hierarchical clustering. Task~1 shows partial tool overlap in the solution space, while Task~2 shows stronger separation, highlighted by the lack of clusters. Log distance was used for Task~2 for legibility which underscores the  solution space size.
    }
    \label{fig:quant}
    \vspace{-2em}
\end{figure}

\parahead{Gulf of Evaluation} First, consider evaluation, or the way in which a user understands how something has changed.

Feedback immediacy scales with how directly a tool exposes the underlying data.
Tools that surface data as a first-class object enable continuous, in-place feedback, while tools that mediate access through code or conversation introduce latency and indirection.
Excel and OpenRefine exemplify this approach by
affording direct data manipulation, such as in Excel's grid-based in-place data editing.
In contrast, Jupyter abstracts data, only making it available through particular views (\eg{} ``.head()'' or ``.info()''), enabling manipulation only through explicit, viscous~\cite{green1989cognitive} commands that must be executed before feedback is available.
ChatGPT abstracts the connection to data even further, connecting only through explicitly requested or automatically selected previews.
None of these strategies led to clearer quality in data cleaning in our study.

However, they do yield differing affective perceptions of the wrangling experience.
Participants noted a diminished sense of control over the data when feedback was insufficiently detailed or punctual.
For example, \pai{7} expressed frustration with ChatGPT:
\qt{I like being able to look at [the data] and kind of get an idea of the overall scope and what is happening within the data set...I didn't really have that ability [with ChatGPT], I relied on giving examples and basically what it was spitting out to me.}
In contrast, more immediate and incremental feedback was perceived as supporting better understanding of the data set and wrangling process.
\plb{3} highlighted the benefits of interactive, incremental feedback execution in Jupyter: \qt{The most helpful aspect of the tool Jupyter was that I could run code interactively and see the output immediately. This made it easier to check results and adjust the code if needed.}
Feedback immediacy shapes not just efficiency but participants' felt understanding of their data.

Meanwhile, some participants (\pai{5}, \pai{9}) expressed positive attitudes toward certain aspects of ChatGPT feedback. They found the step-by-step summaries to be clear, and felt confident with the data wrangling reported. Although this optimism may reflect misplaced trust consistent with automation bias~\cite{skitka1999does}---as their perceived result data quality may not be well aligned with actual result quality, per \figref{fig:quant}---this suggests that AI may create a sense of ease and fluency during the process as it takes over the most tedious work (per \figref{fig:preferences}, the ChatGPT group reported low frustration and mental-demand).
However, the opacity of its underlying operations can lead users to lose actual control over how results are produced.
This may explain why some participants (\pai{3}, \pai{8}) indicated primarily use AI as a co-pilot alongside other tools (\eg{} Excel or Jupyter), assisting in writing code, rather than relying on it as a primary environment for data wrangling.

Transparent and interactive workflows seem to be central to giving a sense of agency and reducing uncertainty.
The experience of control over data depends less on what a tool can do, than on whether its feedback makes its effects legible.

\parahead{Gulf of Execution} Next, consider the gulf of execution which captures the gap between a user's intention and the actions required to achieve those goals within a system.
While these systems exhibit typical gulfs of execution for computer systems, among them was a negotiation between expressiveness and legibility.
For instance, AI tools' opacity can make it difficult for users to anticipate the effects of their actions. Error-prone specifications in code-based environments increase the difficulty of achieving intended outcomes, where small syntax or logic mistakes can lead to failure.

ChatGPT is representative of the opaque end of this spectrum, as its behavior is difficult to predict or control from the user’s perspective. Although natural-language interaction ideally allows users to express complex transformations concisely, this benefit is undermined when the system misinterprets, under-specifies, or over-generalizes users' intent---an issue compounded by the ambiguities of natural language.
As \pai{8} noted: \qt{Sometimes [ChatGPT] goes into some detail that you did not instruct it to do, without taking your own inputs.}
In Jupyter, by contrast, this gulf is closely tied to coding's error-proneness~\cite{green1989cognitive}.
Participants often spent considerable time working on logical or syntax mistakes in code.
While the similar error-proneness exists in Excel and OpenRefine formulas or expressions,
the primary execution gulf of them was manifested in mismatches between users’ expectations and system behavior.
To wit, in OpenRefine, participants (\por{1}, \por{2}, \por{5}, \por{6}, \por{8}) attempted to apply numeric facets to non-numeric columns that appear numeric (\eg{} years).

Another source of execution gaps was feature complexity, where operations exist but their correct use required procedural knowledge that users did not have. In Jupyter, this complexity came from the programming required to translate intentions into valid code, including choosing libraries, writing correct syntax, and managing dependencies.
In Excel, the complexity of advanced features often creates barriers to their correct use, as they have some rigid procedural constraints that users must adhere to. For example, \pxl{1} failed to create a pivot table until he chose a valid data range.
Many participants initially attempted to apply more careful and considerate approaches, but were hindered by the complexity of required operations, giving way to simplified strategies (\eg{} deleting data instead of performing precise transformations).
For example, in Task~2, \pxl{9} created a new column to fill missing values in the ``drive'' column of the car dataset using the most common value. They expected the new column to replace the original one after deletion. Yet, after deleting the original column, the new column produced errors due to a formula dependency. The participant then abandoned this approach, instead removing entire columns containing missing values.

Some features are illegible, as being challenging to discover. For example, \por{1} reported that she failed to find the correct bulk selection place so she: \qt{had to select them one by one, and there were a lot of years, which took a lot of time.}
In the systems examined, the interaction primitives are sufficient to carry out the base wrangling operations (allowing such workarounds), but the more advanced features that would improve efficiency and quality of life were sometimes challenging to identify---a persistent interface design problem~\cite{matejka2013patina}.

Misunderstanding features was also common.
For example, participants (\por{3}, \por{4}) interpreted the facet function as subsetting the data in place, whereas it actually provides a filtered view without modifying the underlying dataset.
In Jupyter, participants conflated views with underlying data.
Excel's literalist data representation made this distinction more explicit, with none of the Excel participants exhibiting this misunderstanding.
Misaligned mental models can require effort to identify and implement correct operations beyond performing the wrangling itself.
More worrying, however, is that this misalignment can result in misplaced confidence.
Designing systems that clearly communicate how actions translate into data transformations, reducing execution gaps and improving both accuracy and user experience, is critical.

\subsection{Errors: What does the system consider to be an error? How are they prevented and handled?}
\label{sec:errors}

Like any tool, errors are pervasive in data wrangling, both in the data and the process of wrangling it. Here we highlight how systems surface errors in each of those categories.

\parahead{Data Errors}
Detecting and correcting data anomalies is central to data wrangling, yet the support provided by the systems, was sometimes limited. Users often had to manually initiate the process of handling anomalies.
Visible help is often confined to simple cues, like empty spreadsheet cells, or none at all, as in misaligned values.

Instead, participants in our study most often pursued data errors manually,
often by writing expressions or invoking specific functions.
Participants consistently expressed dissatisfaction with the current support provided by tools. For instance, \por{5} noted OpenRefine \qt{doesn't support handling missing values in a different way} and \qt{How can we just feel such missing value? It's difficult}. This, in turn, pushed participants to simplistic responses to errors, such as dropping missing values or replacing anomalies with 0, as these were the most accessible options.
Only a few participants reached sophisticated considerations, like handling within group anomalies (\eg{} a salary might not be high on average but might be an outlier for people within a particular profession).
Most participants needed external help (\eg{} AI or docs) to finish, suggesting that tools could offer richer built-in
assistance.

A common approach was to use visualizations, such as scatterplots and histograms, to assess data quality. All tools provide some level of visualization support, but the depth of support varies.
Features such as OpenRefine’s faceting were valued for their ability to easily surface patterns and irregularities.
In contrast, only a few participants (\pxl{1}, \pxl{8}) attempted to use visualizations (\eg{} scatterplot) in Excel for anomaly detection, finding the process tedious and cumbersome.
This suggests the value of integrated and effective visual support to help users detect anomalies more efficiently.
As \por{8} put it: \qt{You can easily see [anomalies] and filter them [using facet], unlike in Python where you have to execute a script, then go to visualize the data.}
\pxl{8} also said he would like to see Excel: \qt{have the data profile and the column quality on the top [of the interface]}, echoing the design of Wrangler~\cite{kandel2011wrangler} and VSCode's Data Wrangler~\cite{vscode-datawrangler}.
Visualizations are a natural way to understand errors~\cite{ruddle2023tasks}, however, they can also mask issues~\cite{mcnutt2020surfacing}, and so they should be integrated with care.

Across groups, participants expressed a strong preference for automated assistance that can suggest or surface potential anomalies for further investigation. For example, participants in the ChatGPT group frequently reported satisfaction with AI support in identifying anomalies and suggesting next steps and appropriate data cleaning strategies---although we note that this form of automation can be highly biased.
We also observed that participants from other groups often turned to ChatGPT for guidance, such as deciding whether to remove suspicious records or seeking alternative wrangling strategies. For instance, \pxl{1} noted they \qt{check[ed] ChatGPT or Google to find certain calculations, and certain practices on how to clean the data.}
Integrating automated assistance (as exemplified by natural language interfaces like Dango~\cite{chen2025dango} or ViseGPT~\cite{zhu2025visegpt}) shows substantial promise in helping users in the potentially unfamiliar wrangling domain, however we stress that like visualizations, these can conceal and propagate errors, and so should warrant careful design and evaluation.

\parahead{Usage Errors}
In addition to errors in data, wranglers are also susceptible to errors in their usage.
In the tools examined, error correction is laborious and fragmented, requiring users to expend substantial cognitive effort to reconcile the tool's output with their intended cleaning goals.

Tools vary in how they respond to errors,
ranging from explicit blocking to implicit signaling.
In Jupyter and Excel, operation errors (\eg{} syntax errors, invalid cell or variable references, or runtime exceptions) are most often detected during program execution and terminate execution with explicit error messages, offering little built-in support for error correction.
For example, Excel provides assistance such as basic formula suggestions and argument hints, but no support for checking usage errors.
In contrast, OpenRefine rarely blocks improper usage; instead, it fails silently, such as by showing that zero records were affected or providing empty visualizations.
ChatGPT, by comparison, lacks an inherent mechanism for detecting or responding to its own errors. Unless users explicitly identify and request corrections, the system does not revise its outputs, allowing errors to remain hidden.
This places a significant burden on error message quality, which, echoing error messages in programming more generally~\cite{becker2019compiler}, is challenging to get right.
Silent failures, in which logical errors are undetected, are a persistent risk.
These include subtle misuse of functions or unintended data transformations that do not trigger explicit error signals---such as the conflation between views and underlying data in Jupyter.

As a result, error handling is largely pushed to the user, typically
relying on manual inspection and iterative verification---highlighting the necessity of manual inspection tools.
For example, \plb{1} and \plb{2} identified and corrected an error only after repeatedly reviewing intermediate outputs and reverting code changes.
Current tools seem to primarily detect errors that violate their underlying execution rules or system constraints, but provide limited support for addressing common cognitive misconceptions during data wrangling, or for anticipating and surfacing likely sources of user misunderstanding.
While extensions to these tools to detect unintended errors exist, such as ExceLint~\cite{barowy2018excelint} for Excel and Retrograde~\cite{harrison2024jupyterlab} for Jupyter, the use of these tools remains limited.

A critical way to deal with errors is to give users a means to manipulate the operation history, such as through undo / redo.
OpenRefine offered the richest support in this regard, allowing users to inspect and roll back to specific previous stages (\eg{} \por{2} utilized the history to revert a flawed transformation).
In contrast, other systems either lack such fine-grained provenance or are not regarded as particularly useful.
For instance, \pxl{1} complained about the difficulty of recovering from changes in Excel: \qt{When you're in Excel, it's just such a challenge to work with the data, because you're, you know, as you delete things, you can't go back to the original state}.
In response to challenges to undo or address issues, participants sometimes resorted to drastic actions.
For example, \pxl{9}, after failing to replace missing values due to unresolved operational errors, deleted all columns containing visible missing values.
Ineffective error feedback can lead users to adopt overly aggressive strategies, preventing them from creating better data wrangling outcomes.

\subsection{Conceptual Structure: How is meaning constructed? How are internal and external incentives balanced?}
\label{sec:Conceptual}
Different tools shape how users construct meaning and approach making sense of data. Conventional wranglers might prioritize breadth or specific tasks,
but burden the user with conforming their needs to the tools' model of task---increasing cognitive load.
In contrast, AI-based tools foreground task context while collapsing intermediate steps, reducing effort but constraining inspection and iterative reasoning.

In Jupyter, wrangling is expressed in code organized in notebook cells, resulting in an incremental sensemaking workflow driven by a composition of abstractions.
This provides high compositional flexibility and supports diverse data cleaning strategies, but requires users to translate high-level goals to general-purpose programming constructs.
Notebooks require users to explicitly invoke inspection methods (\eg{} .head() or .isnull()), and the abstractions users apply shape which data states become available for inspection. As a result, subsequent actions are often driven by the specific inspection output they get.
Identifying a visible anomaly after applying .head() may lead to operations such as dropping rows.
The sense-making is a reactive process driven by abstraction and its output.

In contrast, Excel and OpenRefine employ domain-specific conceptual models that align closely with the data's layout.  Excel utilizes spatial compositionality, where cells serve as both data containers and sites of transformation. This WYSIWYG structure enables direct manipulation through localized expressions. OpenRefine narrows this proposition further by embedding common operations within a column-level schema, framing data wrangling as a linear ``detect-and-repair'' pipeline centered on facets (providing an interactive visualization of column distributions, allowing users to instantly spot anomalies and filter data through direct engagement with these visual clusters) to identify underlying patterns and execute batch transformations.
In both cases, the grid centered display
tended to beget opportunistic (rather than strategic) cleaning, wherein users address data quality issues iteratively as they are discovered rather than systematically. For example, when checking for anomalies, \pxl{7} did not follow a systematic plan (\eg{} inspecting numerical or categorical columns in sequence), but instead first focused on a column with a large number of missing values at first glance.
This tool-driven behavior often led participants to prioritize issues that were easily addressable via prominent functions (\eg{} clustering or filtering) over more critical task objectives.

\begin{figure}
    \centering
\includegraphics[width=\linewidth]{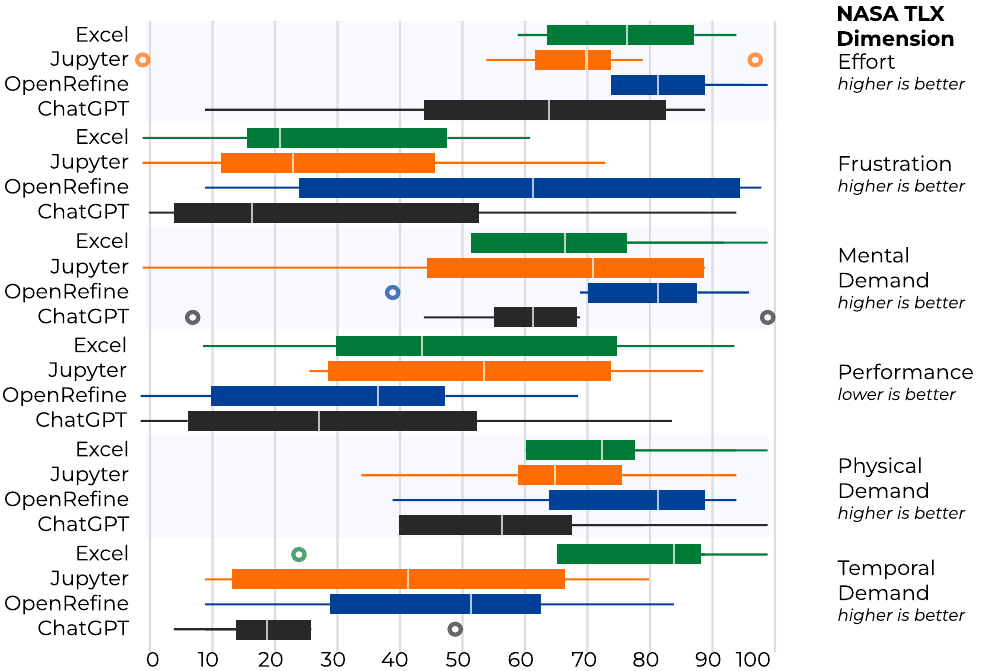}
    \caption{Participants held a range of opinions about the tools they used, with no tool consistently faring better than any other.
}
    \label{fig:preferences}
    \vspace{-2em}
\end{figure}

ChatGPT participants' sensemaking was largely constructed in a passive and task-driven manner. Participants relied on the AI to interpret data quality issues and propose cleaning actions, often adopting these suggestions directly. As a result, participants were less likely to explore alternative strategies or engage in iterative reasoning. This interaction style led participants to expect ``one-shot'' solutions, specifying requirements upfront rather than refining them over time. This appears to reduce effort compared to other tools, as participants reported lower frustration in \figref{fig:preferences}. As in \secref{sec:interaction}, this reliance reflects overtrust in the AI---automation bias.
When the AI's reasoning diverged from user intent, the lack of inspectable intermediate steps hindered understanding and correction. This is also reflected in \secref{sec:quant}), where the ChatGPT group did not reliably produce satisfactory results.
While fully automated approaches still remain ill-suited for these detail-oriented tasks (requiring iterative reasoning  and inspection), task-oriented AI-driven strategies are a promising direction for data wrangling.

No single tool seemed to completely support users in forming a clear sense of their target outcomes, as reflected in high uncertainty on how to improve results further
and frequent reliance on external resources for guidance (especially outside the ChatGPT group).
This reveals a gap in how current systems support sensemaking during data wrangling, integrating AI-based guidance or more structured task support could reduce unnecessary exploration and shift between multiple tools.

\subsection{Notation: How are the different notations related?}
\label{sec:notation}

Data wrangling systems manage their complexity by structuring different notational elements---such as operations, outputs, and error messages---into distinct but interrelated representations.
Effective systems balance operations and outputs that are both distinguishable and tightly integrated at the notational level, supporting stable mental models, while a notational system that poorly structures state boundaries could lead to error-proneness~\cite{green1989cognitive}.
For example, in Jupyter, code inputs and outputs are co-located while remaining visually distinct within cells, allowing users to directly associate actions with results without navigating across interface regions. It also provides secondary notation to help users manage state and track progress in order to address the challenges of increasingly long workflows in notebook environments.
Similarly, OpenRefine embeds commonly used transformations in  column headers, putting relevant operations in close proximity to the corresponding data.
This tight coupling supports quick reuse and helps users develop lightweight, repeatable workflows.

Excel separates operations and outputs across different interface regions and notations: the grid versus the ribbon, the displayed value versus the formula generating it. Participants often had to switch between views to inspect results and locate controls, introducing friction and fragmenting their workflow.
While prior work shows that such interleaving can support ongoing sensemaking~\cite{bartram2021untidy}, it can also introduce ambiguity when not properly managed. For example, \pxl{1} submitted incorrect draft sheets instead of the output data, and \pxl{5} and others left not fully cleaned intermediate transformation steps embedded in the result data, as they were cut off by the view. This increases the occurrence of easily overlooked errors.

In contrast with more explicit and formal notations, ChatGPT relies on a more informal, ad hoc notational system.
Interaction is expressed through a single natural language interface, where instructions, operations, outputs, and explanations become entangled in the same conversational stream.
While this lowers the operation barrier, it also blurs distinctions between different stages of interaction, requiring users to interpret responses to identify relevant information.
In addition, its long, report-like responses make it harder for users to retain and trace prior operations.
Further, these responses are inconsistent and may change over time, making it challenging to develop familiarity or reliance on specific forms and formats.

Effective design here requires balancing distinguishability and integration. When notations related to recurring workflow are fragmented across representations, interaction becomes discontinuous. When different structures are compressed into a single representation, their distinctions are obscured, making workflows harder to trace and error-prone.
Both cases hinder users’ ability to construct, maintain, and reason.
Introducing a DSL as a target notation is an established pattern in this domain~\cite{chen2025dango, gorinova2016end, kandel2011wrangler}, suggesting that separating transformation steps from the data while keeping them within a unified workflow representation can help mitigate this tension.

\subsection{Complexity: How does the system structure complexity and what level of detail is required?}
\label{sec:complexity}

Data wrangling involves both executing transformations and managing higher-level strategies. While automated assistance (\eg{} built-in functions) abstracts low-level complexity, excessive automation in workflow management can obscure reasoning and obscure user agency.

Domain-specific tools can streamline the interaction loop by modeling the intended task.
To wit, OpenRefine users highlighted facets and common transformations as being particularly helpful: \qt{I love the transformation part} (\por{6}).
Similarly, Excel supports aggregation via pivot tables and   explication of data relationships via formulas, although these can be challenging to author correctly.
Without these features, users must manually specify operations and repeatedly update data, increasing both effort and the risk of inconsistencies---a persistent concern in spreadsheet-based data work~\cite{bartram2021untidy}.
Although notably, these tools can not do everything.
For example, \por{5} noted limited support for handling missing values, while \pxl{1}, \pxl{8} and \pxl{10} all pointed out that checking data quality and distribution are still tedious processes in Excel compared to visualization-focused tools like Power BI.
Some of this dissatisfaction stemmed from misunderstandings of available functions.
While specifying to a domain necessitates losing some generality, effective design negotiates that trade off to address these and other specific concerns.

In contrast, wrangling usage of Jupyter is centered on external libraries (\eg{} pandas).
This rearrangement of complexity supports richer expressivity and compositionality,
however this forces users to work within the framework of a library, potentially creating greater cognitive load.
This flexibility comes at the price of isolation: users of explicitly wrangling-focused tools tended to just use the tools, while Jupyter users used more external support (\eg{} AI assistants).

Reducing complexity is not a simple matter of increasing the degree of automation.
For example, \pxl{4} was frustrated by Excel automatically changing data types (\eg{} converting numbers to dates), violating expectations and disrupting the workflow---highlighting how overeager(or impolite~\cite{whitworth2005polite}) tools can impede usage.
In another direction, ChatGPT allows users to specify desired outcomes declaratively, reducing the need for explicit procedural specification. However, this shifts complexity to prompt formulation and interpretation, while limiting users’ ability to inspect and control intermediate steps, sometimes leaving them at a loss for what to do next.
To this end, many participants reported in post-task surveys they submitted their results not because they were satisfied, but because they did not know how to further improve them ($5.2\pm0.8$ on a 7-point scale).
Further echoing Whitworth\cite{whitworth2005polite}, effective assistance should support users without overriding or replacing their control. More specifically, providing scaffolding (\eg{} checklists or runbooks) seems especially valuable, as it supports reasoning about progress and next steps.

\subsection{Customizability: Once a program exists in the system, how can it be extended and modified?}
\label{sec:Customizability}

Customization needs (here meaning mechanisms for post-construction reuse) range from ad hoc operations to reusable transformations~\cite{gridlets, papermill} that can be applied across different parts of the data (\eg{} applying the same missing-value handling to multiple columns). The tools we studied constrain customization within predefined frameworks, and their design trade-offs shape whether users formalize reuse (\eg{} functions in notebooks) or rely on repeated, unstructured ad hoc operations (\eg{} copy and paste).

In programmatic environments such as Jupyter, participants frequently highlighted the advantages of functions and  well-defined external libraries (\eg{} pandas, matplotlib) that support composing reusable building blocks rather than ad hoc procedures. For example, we observed that \plb{2} wrote a function to check outliers across columns, although most users did not use functions.
Code-centric environments afford formalization of recurring transformations into reusable abstractions, although the use of abstractions like functions is reasonably uncommon in computational notebooks~\cite{rule2018exploration}. For those that do use them, this modularization can help improve readability and maintainability at the price of some construct complexity. Exploration of code-based wrangling practices outside of computational notebooks may further be revealing about the relationship between reuse-focused abstraction and wrangling.

Many wrangling tasks can be performed via simple manual actions in direct manipulation-centered tools.
While Excel supports reuse via formulas and macros, leveraging these features often requires a shift to a separate programming layer. In practice, users tend to write formulas directly in cells without leaving the interface. The tabular layout also makes it easy to see both the data and the results of operations at a glance~\cite{bartram2021untidy}, making ad hoc customization easy and accessible in Excel---which might also be tedious and error-prone (\secref{sec:notation}).
OpenRefine reflects a tension between flexible, in-place interaction and the need for structured, reusable workflows. Although it supports direct manipulation, defining transformations through expressions requires switching to a separate interface, and it provides expression history as a lightweight form of reuse. Some systems (\eg{} Trifacta~\cite{kandel2011wrangler} or Buckaroo~\cite{rezig26buckaroo}) provide script export to better support reuse within primarily direct-manipulation workflows.
ChatGPT provides little support for structured reuse, mainly relying on prompting and integration with external tools.
This echoes a larger trend towards irreproducibility in  ML-driven workflows~\cite{kapoor2023leakage}.

Paradigms shape reuse and customization.
Direct-interaction environments encourage repeated, in-place actions that are easy to apply but prone to errors, while programmatic systems encourage structured reuse that reduces repetition but introduces additional complexity.
This tension helps explain why reuse remains challenging in practice.
Domain-specific tools that integrate direct manipulation with structured operation management may offer a promising direction.

\section{Limitations}
Our study had several limitations.
Key among them is that the approaches used in the context of a (remote) lab study may differ from the practices that users employ in more everyday data cleaning, which is an inherent limitation of lab studies. While fixed tasks supported a more straightforward comparison, future work could use in-the-wild methods, such as interviews and logging, or have participants bring their own data to better reflect real-world practices.
As participants could use external resources, their results may also be shaped by support beyond the assigned tool condition. Our findings should be interpreted as primary tool environments rather than isolated treatments.
Similarly, there may be effects related to our hosting of tools in a VM, rather than natively, or our use of JupyterLite rather than JupyterLab (although those differences between the two are minor).
We used datasets that required limited context and focused on cleaning tasks, though such context can shape effective use~\cite{bartram2021untidy} and trust~\cite{sultanum2024data}, and the observed pattern may differ in integration or harmonization tasks.
Similarly, the specifics of our tasks may have shaped user strategies, although few participants accounted for all the required considerations.
Future work should examine other wrangling tasks and different data scales.
Our study was fundamentally exploratory in nature, with a primary focus on usage patterns situated against broader performance trends. We therefore drew on methods (\eg{} Jaccard, TDoPs) that support coarse, qualitative evaluation. While this approach may miss some fine-grained patterns, it appears to capture meaningful patterns broadly. Measuring qualities of cleaned data from human-in-the-loop contexts poses an interesting challenge; future work could consider refinement methods such as multi-set comparison, and explore richer measures that better capture semantic differences.
Similarly, our between-subjects design limits direct inter-tool comparison. Although useful, a within-subjects design would have impractically extended sessions and shifted participants into comparative evaluation rather than reflecting everyday tool use.
While we tried to examine a representative cross-section of data-wrangling paradigms, we omitted node-and-wire systems (\eg{} Alteryx) due to price issues and concerns about our ability to recruit participants, though we expect our findings would be similar.
We unintentionally did not collect demographic data (\eg{} gender) because of our transition from Prolific (where it is provided automatically) to freelancer platforms (where it is not).
We used relatively harsh language in task instructions (to address concerns about AI use in crowd-work), which may have affected participant behavior.
Lastly, our study was conducted at a particular moment in time, and as society's relationship with technology changes (\eg{} with AI) the saliency (or even content) of these dimensions may shift.

\definecolor{secA}{HTML}{DCDCDC}
\definecolor{secB}{HTML}{DCDCDC}
\definecolor{secC}{HTML}{DCDCDC}

\newcommand{\finding}[1]{\textbf{#1}.}
\begin{table*}[t]
    \caption{
Here we highlight several key tensions which shaped participants’ wrangling practices identified through our TDoPS analysis.
}
    \label{tab:tdops-synthesis}
\rowcolors{1}{gray!15}{white}
    \setlength{\aboverulesep}{0pt}
    \setlength{\belowrulesep}{0pt}
    \setlength{\extrarowheight}{2pt}
    \begin{tabular}{p{0.98\linewidth}}
        \toprule
\finding{Visibility vs. Abstraction}
        Visibility-driven interfaces can make salient data issues easy to inspect, encouraging opportunistic cleaning.
        In contrast, abstraction-centered workflows support more systematic and compositional reasoning at the cost of greater cognitive and procedural effort.                                                                                                                                                   \\
\finding{Predictability vs. Simplicity} Natural language UIs make complex transformations easy to specify, but introduce ambiguity and make their behavior harder to predict; whereas more explicit specifications offer greater control at the cost of additional procedural effort and error-proneness. \\
\finding{Big vs. Small Steps} One-shot automation simplifies multi-step transformations, but can obscure intermediate states, scope, and downstream effects---reducing users' ability to validate individual operations, make fine-grained adjustments, and rollback incrementally from errors.           \\
\finding{Error Visibility vs. Attentional Capture}
        Explicit error signals can make some anomalies more visible, but doing so can narrow attention to predefined issue types and draw attention away from other issues and processes.                                                                                                                         \\
\finding{Entangled Integration vs. Fragmented Splits} Separating data, transformations, and outputs improves distinguishability but can increase navigation and fragment interaction, while tighter integration supports more continuous work but risks entangling operations and outputs.                \\

        \finding{Well-trodden Pathways vs. Unguided Flexibility} Domain-specific tools simplify common wrangling operations but provide limited support for less typical transformations, in contrast general-purpose tools offer flexibility at the cost of higher procedural and programming demands.           \\
\finding{Structured Reuse vs. Ad Hoc Repetition} Lightweight, in-place interactions can make transformations easy to apply but often encourage ad hoc repetition, while structured abstractions support reuse and maintainability at the cost of additional authoring and management complexity.
\\
        \bottomrule
    \end{tabular}

    \vspace{-1em}
\end{table*}

\section{Discussion}

This work seeks to characterize the usability of data wranglers via consideration of their technical dimensions~\cite{jakubovicEP23}.
Data wrangling is an inherently challenging task, requiring contextual judgment and iterative decision-making, rather than pursuit of a single correct solution.
These challenges manifested as a range of design trade-offs that shape the wrangling process. Table~\ref{tab:tdops-synthesis} consolidates the comparisons developed throughout our TDoPS analysis.
We stress that the identified dimensions are non-normative: there are lots of good ways to build an effective wrangler. Instead we suggest that consideration of these dimensions may lead to more intentional designs.
We reflect on this analysis and note future work.

\parahead{On AI}
Our findings point to potential risks associated with overreliance on AI, including misalignment between interfaces and users’ mental models, as well as the introduction of silent errors. While AI-integrated wrangling tools become increasingly common\cite{petricek2022ai,chen2025dango,li2024towards}, how to use them effectively to improve usability, user confidence, and explainability remains an open question~\cite{passi2022overreliance}.
Although AI-driven interfaces can reduce cognitive effort, lower the learning curve, and improve efficiency---as reflected in our study, where participants in the ChatGPT condition generally spent less time (see \figref{fig:timing}), they may also come at the cost of a deeper understanding of the underlying data wrangling process.
Intermediate states may be less visible and interactions with natural language can be difficult to interpret or predict, especially when model behavior is inconsistent. This can lead to frustration and reduced trust.

At the same time, AI use crossed tool-condition boundaries in our study. Although ChatGPT was assigned as one primary tool condition, participants in every condition were allowed to consult external resources, and those who were in non-AI conditions usually turned to AI (such as ChatGPT and Claude) when they needed help writing formulas or expressions--- which helped participants translate their wrangling intent into executable operations within that tool. In this sense, AI functioned as an ambient support layer around conventional wrangling environments, especially when users encountered notational or procedural barriers.
Echoing its use in coding, an effective role for AI may be as a co-pilot integrated into existing tools, rather than a primary interface. The way that participants used ChatGPT in this study might be characterized as \emph{vibe-wrangling} which naturally inherits all the challenges and errors latent to that usage model~\cite{sarkar2025vibecoding}. Per the cognitive disengagement~\cite{lee2025impact} that typically comes from heavy AI usage, we highlight that designing tools such that domain focus is retained as a central research challenge.
For instance, AI might be used to improve error messaging to offer more effective diagnostic guidance and suggestions, assist code generation for basic transformation\cite{li2024towards}, or be used as a means to operate a secondary source of truth---such as a DSL, as in Dango~\cite{chen2025dango}.

\parahead{Tools shape how users wrangle}
Tool affordances can steer user strategies, but they do not determine outcomes. Within the scope of our study, there is no single tool that consistently outperformed others, nor is there a one-way mapping between tools and technique results. Instead, tools primarily influence how participants approached sensemaking, sometimes at the expense of more appropriate or effective data wrangling strategies.
In Jupyter, participants relied heavily on data previews (\eg{} summary statistics and table views) to identify visible anomalies.
Whereas in OpenRefine, participants frequently focused on rapid type conversions and were drawn to prominent features such as clustering and reconciliation.
This suggests a form of tool-induced sensemaking, in which users implicitly redefine their tasks in terms of the operations the system supports.
Tools do not push users toward the same answer, because they do not support the same way of seeing.

The tools we examined spanned a range of specificity in data wrangling.
General-purpose tools (Excel and Jupyter) provide broad flexibility and powerful functionality, while systems like OpenRefine emphasize specific transformation capabilities.
Again, users’ wrangling processes are not solely driven by their initial intentions. Instead, the available functionalities and representations of the tool actively shape how users formulate problems and construct solutions.
In the tools we examined, there was sometimes a gap between high-level intent and actionable steps, often forcing users to bridge this gap manually via trial-and-error. Consequently, rather than following a planned, goal-directed process, users were guided by the logic and visibility of the tool itself.

This is an innate challenge of interface design~\cite{brooks1987nosilverbullet}: how to surface the inherent complexity of the problem versus the accidental complexity of the interface.
There is, of course, no one answer and any design will be a negotiation between these factors; a challenge which is multiplied by the tensions between domain-specific wrangling tools versus general-purpose tools that can be used for wrangling.
One promising direction is to incorporate use-case-oriented scaffolding, such as run books,
or interactive guidance mechanisms, like linters~\cite{barowy2018excelint, harrison2024jupyterlab}---potentially augmented by AI.
These approaches could help users identify relevant operations, preview outcomes, and explore alternatives without relying on unguided trial and error, thereby improving system usability---this does not diminish the value of exploration, as trial and error can itself support learning about both the interface and the domain.

The interfaces studied parse this design space in different ways, so an interesting approach might (further) explore interface paradigm mashups.
For instance, notebooks are often used with AI assistants~\cite{mcnutt2023design}. Further combinations are worth exploring, such as intersecting the visual primacy of spreadsheets with the abstractions of notebooks.

\section*{Acknowledgments}
We thank our participants for sharing their time and insights with us, as well as our reviewers for their thoughtful and engaged commentary. We thank Joseph Hellerstein for insightful feedback on this work. We also thank Baolu Yu,  and Will Jackson for piloting this study.
This work was supported in part by the NSF award \#2402719.

\bibliographystyle{IEEEtran}
\bibliography{wrangle}

\clearpage{}
\onecolumn{}
\appendix{}

\begin{figure}[t]
    \centering
    \includegraphics[width=0.5\linewidth]{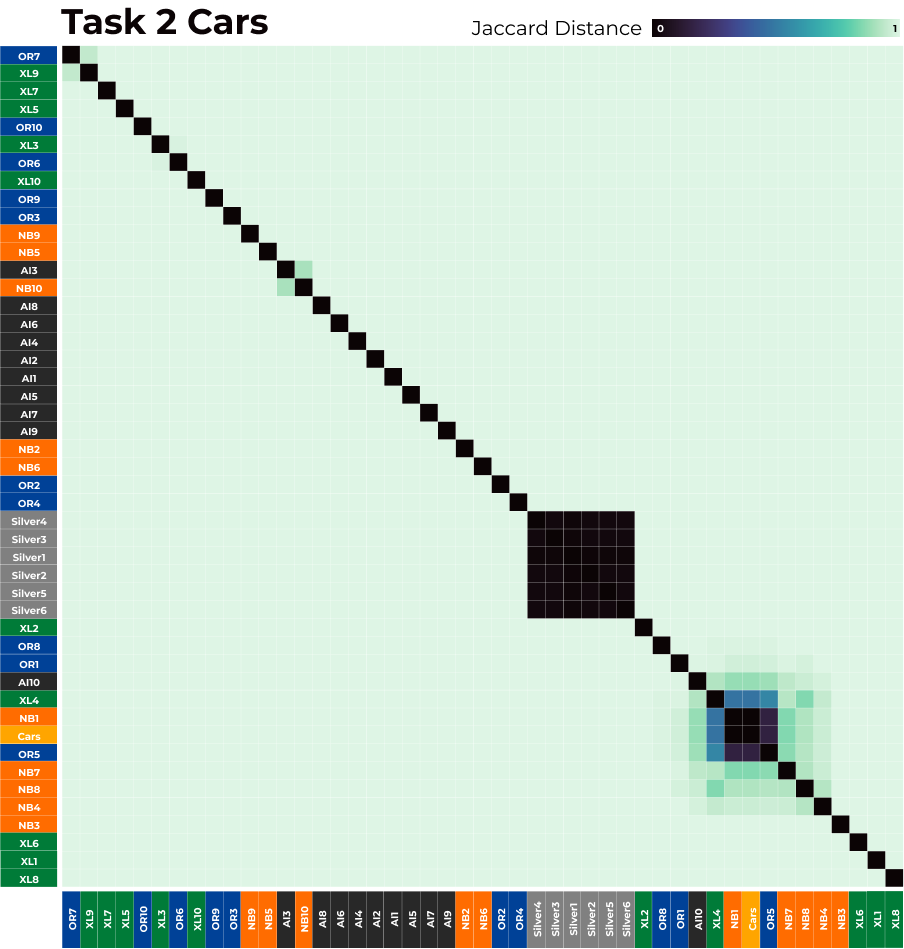}
    \caption{A version of \autoref{fig:quant} with task 2 presented using a linear color map. Compared with the one in the main body of the paper this one makes it significantly more challenging to view the clusters present in the data. This highlights the high dis-similarity between participant generated datasets for the second task as well as the relatively high similarity between the pre-prepared silver tables. }
    \label{fig:quant-logged}
\end{figure}

\begin{figure}
    \centering
    \includegraphics[width=\linewidth]{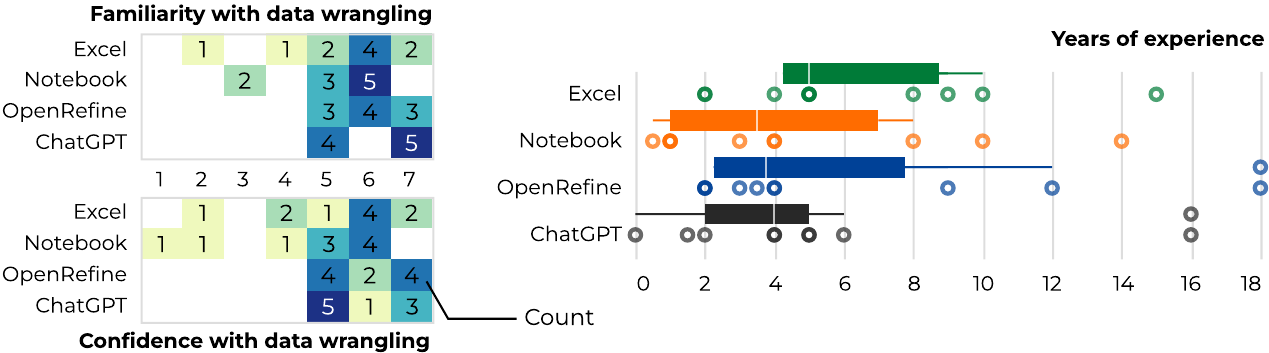}
    \caption{Participant familiarity, confidence, and experience by tool. }
    \label{fig:exp-wrangling}
\end{figure}

\section{Appendix}

Here we include additional materials that did not fit within the main body of the paper.
Additional study materials for the study are available at \osf{}, including analysis code, datasets, stimuli code, analysis spreadsheet for TDoPS.
An example of the JupyterLab version of the study can be found at \url{https://wrangler-tdops-study.netlify.app/a-notebook}. Among the four tools checked, this is the only one that does not require an active connection to a server, and so we provide it as a standalone link. 
For convenience, we also provide this link, \url{https://github.com/mcnuttandrew/wrangler-tdops-study} to our study and analysis code. In \figref{fig:quant-logged} we show a variation on the quantitative results figure from the main paper to highlight a design decision. 
In \figref{fig:exp-wrangling} we show the distribution of self-described experience questions.

\parahead{Procedural Criteria for TDoPS Coding} An observation was coded as evidence of a TDoPS dimension when a participant either demonstrated relevant interaction behavior or explicitly described an experience reflecting that dimension, such as satisfaction, difficulty, or friction. For example, difficulty locating or correctly applying a feature was coded as evidence of the gulf of execution or feature illegibility, while comments about an inability to inspect intermediate results were coded as evidence of the gulf of evaluation.

\parahead{Experimental Configuration Details} Here we describe the specific versions of each of the tools used. 
Participants in the ChatGPT condition used the standard ChatGPT web interface in Chrome,
with the model ``ChatGPT-5.2'' at the time of the study. The interface
allowed participants to upload CSV files, interact with the model through natural-language prompts, and download cleaned outputs. OpenRefine condition used OpenRefine 3.10.0. Participants in the Excel condition used Excel for the web. These three conditions were accessed through a Windows VM embedded in ReVISit via noVNC, in order to support remote deployment and consistent instrumentation while addressing constraints around licensing, installation.
Participants in the Jupyter condition used JupyterLite~\cite{jupyterLite}, a browser-based notebook environment that supported the packages needed for our tasks, including pandas. This condition was deployed through Netlify and Render.

Some deployed configurations differ from common desktop or server-backed environments---for example, Excel for the web rather than desktop Excel, and JupyterLite rather than a hosted JupyterLab instance---due to remote experimentation constraints. 
However, each configuration preserved the core functionality and interaction patterns that users needed to complete our study tasks, serving as a representative implementation of the broader tool paradigm examined.

\subsection{Task prompts}

\subsubsection{Task 1: Wrangling Game Rating Dataset}

{\ttfamily

You are a data analyst at a gaming media outlet, and you need to analyze how User Scores have trended by Genre since the year 2000. The raw data is currently too messy for the pipeline. Please clean it using the assigned tool.

In this task, you will clean the provided dataset using the assigned tool. Your goal is to prepare a dataset in which all fields relevant to the analysis are clean, consistent, and ready for computing annual average scores. Please select appropriate wrangling strategies and correct abnormal values where necessary.
Your Goal:

Filter: You are only interested in the modern era. Please exclude any records released prior to 2000.

Standardize: Ensure 'User\_Score' is numeric and ready for aggregation.

Quality Control: We need a high-quality dataset. Trends will be invisible if the data contains abnormalities such as missing values, type errors, unreasonable outliers, or any other inconsistencies. So please choose reasonable strategies to deal with those abnormalities.
Notification:

Be Conservative: Do not aggressively delete rows. If data is messy, try to save the record by imputing values based on reasonable logic. Only delete rows if the data is unrecoverable.

Context Matters: Be careful when flagging outliers. A value that looks strange globally might be normal within its specific subgroup. Think of housing prices as an example: a price that is an outlier for the whole country might be normal for a city like New York. Verifying anomalies within their groups before removing them is valuable.
Estimated Time

Please try to complete this task within 10 minutes. Do your best, but avoid spending more than 10 minutes on it.
Tutorial of Submission (Please make sure you watch this video carefully)

Important Tips Before You Proceed
Connect to the VNC Server

1.Click "Connect" and enter the VNC Passcode(You can find it in the sidebar)

2.Click the toolbar, then click the button labeled A. Select the last option to send "Ctrl + Alt + Del", and then enter the Server PIN (which can also be found in the sidebar).

3.Please do not attempt to access any other applications on the server we provided. You may only use the ChatGPT window available on this server. If you need external resources (such as Google searches), please use your own web browser.

4.Please ensure that you save the cleaned dataset from ChatGPT correctly as demonstrated. For example, once you are satisfied with the results provided by ChatGPT, you may use a prompt such as: “Please return the cleaned version of this CSV file for downloading.”

5.Please make sure you open the correct file as required. For Task 1, use "game.csv", and for Task 2, use "car.csv".

6.Please open a new chat to demonstrate Task 2, and do not delete any chat history created during the study.

Any violation of the above instructions, or failure to complete the study as required, will result in the study being considered incomplete, and NO COMPENSATION will be provided.

}

\subsubsection{Task 2: Wrangling Used Cars Dataset}

{\ttfamily

You are a data analyst at a used car trading platform, and you need to prepare a dataset to train a "Car Value Retention Model" for various manufacturers. The raw data is currently too messy for the pipeline. Please clean it using the assigned tool.

In this task, you will clean the provided dataset using the assigned tool. Your goal is to prepare a dataset in which all fields relevant to the analysis are clean, consistent, and ready for modeling. Please select appropriate wrangling strategies and correct abnormal values where necessary.
Your Goal:

Standardize Data Types: Identify the columns valuable for building the model and ensure they are assigned the correct data types. For instance, ensure numerical values are consistently numeric and descriptive labels are categorical.

Clean Anomalies: Identify and resolve data quality issues such as null values, type mismatches, outliers, or any other inconsistencies. Please apply reasonable strategies to correct these abnormalities.

Remove irrelevant columns: To ensure that the cleaned dataset is suitable for modeling, you should remove any columns that are not informative or appropriate for the predictive task. For example, long free-text descriptive fields or identifiers that do not contribute meaningful features are generally not useful for modeling and should be excluded.
Notifications:

Be Conservative: Avoid aggressive data records deletion. Crucially, ensure that no manufacturer is completely removed from the dataset. If the data is messy, try to salvage the record by imputing values based on reasonable logic. Only delete rows if the data is unrecoverable or irrelevant.

Context Matters: Exercise caution when flagging outliers. A value that appears to be a global outlier may be normal within its specific subgroup. For example, a price of \$250,000 might be an error for a "Ford," but perfectly valid for a "Ferrari."
Estimated Time

Approximately 15 - 25 minutes. But there is no strict time limit for this task, work at your own pace and submit your results once you feel the task is complete. Please do your best to wrangle the data and identify any anomalies. Remember that your goal is to produce a clean dataset that is suitable for modeling.

}

\subsection{Silver Tables Construction}

We construct multiple silver-standard datasets for each task to serve as reference baselines for evaluating participants’ data wrangling outputs. Rather than defining ground truth, our goal is to represent a set of high-quality, task-aligned cleaning strategies that reflect realistic analytical practices. They are designed to:
\begin{itemize}
    \item Align closely with task requirements, ensuring that key instructions (\eg{} filtering data ranges, standardizing data types, and considering outliers in context) are reflected.
    \item Capture meaningful variation in reasonable data cleaning strategies, particularly in how missing values are handled
\end{itemize}

While some strategies (\eg{} dropping missing values) may appear to conflict with task recommendations (such as encouraging conservative deletion), they are included because such decisions are inherently subjective in practice. Determining whether a record is recoverable often depends on the analyst's judgment, and row deletion remains a widely used and acceptable approach when data quality is deemed insufficient. 
These silver tables construction strategies were discussed among the research team and considered to be reasonable reference points for the analysis, while we acknowledged that other valid silver tables could also be constructed beyond this set.
Overall, we treat these silver tables as baselines of good-quality outputs that align with task requirements reasonably, rather than strict optima.

\vspace{1em}

\subsubsection{Silver Tables Generation Process for Game Dataset}

\paragraph{Global Preprocessing}
All silver tables share the following preprocessing steps:
\begin{itemize}
    \item Remove games released before 2000
    \item Standardize \texttt{User\_Score}:
    \begin{itemize}
        \item Convert non-numeric values such as\texttt{``tbd''} to missing values (\texttt{NaN})
        \item Cast the column to a numeric type
    \end{itemize}
\end{itemize}

\paragraph{Silver Table Variants}
All variants detect outliers within groups defined by the \texttt{Genre} column using the IQR method, so unusually high or low values are judged within their genre rather than across the entire dataset. The variants differ only in how they handle missing values in \texttt{User\_Score}.

\begin{itemize}
    \item \textbf{S1 (Deletion):} Drop rows with missing \texttt{User\_Score}, then remove outliers.
    
    \item \textbf{S2 (Constant Imputation):} Fill missing values with 0, then remove outliers.
    
    \item \textbf{S3 (Global Median):} Fill missing values with the global median, then remove outliers.
    
    \item \textbf{S4 (Group Median):} Fill missing values with the median within each Genre, then remove outliers.
    
    \item \textbf{S5 (Global Mean):} Fill missing values with the global mean, then remove outliers.
    
    \item \textbf{S6 (Group Mean):} Fill missing values with the mean within each Genre, then remove outliers.
\end{itemize}

\vspace{1em}

\subsubsection{Silver Tables Generation Process for Car Dataset}

We note that the decision space for this task was relatively large. Decisions such as which columns to remove and how aggressively to handle missing or anomalous values often vary based on analyst judgment and prior experience. Thus, we constructed the silver tables as analytically plausible references, following common data-cleaning practices and conventional choices in data analysis while aligning them as closely as possible with the task requirements.

\paragraph{Global Preprocessing}
All silver tables apply the following preprocessing steps:
\begin{itemize}
    \item Remove irrelevant columns (e.g., \texttt{id}, \texttt{VIN}, \texttt{description})
    \item Drop rows with missing values in the target variable (\texttt{price})
    \item Convert numeric columns (\texttt{price}, \texttt{year}, \texttt{odometer}) to numeric type
    \item Replace missing values in categorical columns with \texttt{``unknown''}
\end{itemize}

\paragraph{Silver Table Variants}

All variants use the IQR method to detect outliers within each \texttt{manufacturer} group across numeric columns, reflecting that vehicle attributes often vary substantially by brand. The variants differ in how they handle missing values in numeric columns:

\begin{itemize}
    \item \textbf{S1 (Deletion):} Drop rows with missing values in numeric columns, then remove outliers.
    
    \item \textbf{S2 (Constant Imputation):} Fill missing values in numeric columns with 0, then remove outliers.
    
    \item \textbf{S3 (Global Median):} Fill missing values with global medians, then remove outliers.
    
    \item \textbf{S4 (Global Mean):} Fill missing values with global means, then remove outliers.
    
    \item \textbf{S5 (Group Median):} Fill missing values with medians within each manufacturer group, then remove outliers.
    
    \item \textbf{S6 (Group Mean):} Fill missing values with means within each manufacturer group, then remove outliers.
\end{itemize}

\subsection{Adoptability: How does the system facilitate or obstruct adoption by both individuals and communities?}

\emph{Here we offer our analysis of the adoptability of different systems, but emphasize that this is based on general observations rather than from evidence generated in our study which has relatively limited bearing on this longer time scale component}

Adoptability considers not merely whether a tool is easy to learn or performs well on specific tasks (although these are important), but how it fits into the broader context in which data work takes place.
This echoes Meyerovich \etal{}~\cite{meyerovich2013empirical} who find (in studying programming languages) that adoption is shaped by the usage ecosystem, rather than merely technical features alone.
Our study primarily focuses on short-term wrangling practice by users with non-trivial expertise rather than long-term adoption or approachability by novices,  so our observations here are necessarily tentative.

Echoing our earlier analysis (\secref{sec:complexity}), we suggest that complexity profile is likely connected with broader issues of learnability. For instance, Jupyter offers rich, powerful tools which can be challenging to learn, whereas Excel has a simpler on-ramp---although getting to the higher levels of proficiency requires non-trivial efforts.
These differences in complexity ceilings shape not just individual learning trajectories but also the kinds of community support that develop around each tool.

Community resources play a significant role in mediating these learnability barriers. 
For instance, Excel is supported by a mature ecosystem of users, documentation, integrations, and even formalized sports coverage~\cite{esports} that together provide a range of ways to learn to use it effectively. 
While the economic investment from major corporations clearly helps sustain such, it is not uniquely determining, as open-source efforts like Jupyter can also thrive---although this is may in part be driven by philanthropic investment. 
OpenRefine illustrates the risks of ecosystem fragility. Following Google's discontinuation of active support in 2012, the tool transitioned to a community-driven development model---a shift that may have constrained its visibility and adoption trajectory, even as its core functionality remained strong. 

Paralleling this form of investment is cultural entrenchment,  where a tool might be dominant simply because it is the most commonly understood for that task, such as how Googling is synonymous with searching the internet.
For instance, \pxl{1} avoids Excel but still acknowledges its entrenched position in data analysis, observing that \qt{it's been around forever}.

Lastly, one sociotechnical element governed by design choices is how different tools resolve functionality barriers.
Some allow for extensions (prompting the library style ecosystem seen in Jupyter), while others prompt changing between tools (prompting tool chain style ecosystems as with ChatGPT).
Connecting with the previous section, these strategies are associated with the level of generality in the tool. 
Users navigate these trade-offs by combining multiple tools. 
For example, \pxl{3} described using Excel to handle anomalies while relying on Python for more targeted analysis, while ChatGPT was widely used alongside other tools in our study (such as for developing Excel formulas). 
Multi-tool adoption suggests that, rather than designing individual, comprehensive tools, there may be value in designing for interoperability and smooth transitions between complementary systems.

Adoptability then is not a fixed tool property but an emergent outcome of a range of different factors including its complexity profile, ecosystem maturity, and cultural position. This suggests that investments in interoperability, community infrastructure, and gradual learning pathways may be more critical than usability improvements.

\end{document}